\documentclass[prl, aps, superscriptaddress, twocolumn, showpacs, nofootinbib]{revtex4-2}

\usepackage{placeins}
\usepackage{amsfonts, amsmath,amssymb,amsthm,bm}
\usepackage{mathrsfs}
\usepackage{graphicx}
\usepackage[usenames,dvipsnames]{color}
\usepackage{palatino}
\usepackage{fancyhdr}
\usepackage{mathrsfs}
\usepackage{multirow}
\usepackage{hyperref}
\usepackage{bigints}
\usepackage{soul}
\usepackage{bbold}
\usepackage{float}

\usepackage{relsize}

\usepackage{framed}
\definecolor{shadecolor}{rgb}{.8,.8,.9}
\definecolor{gr}{rgb}{0,.5,0}

\usepackage{braket}

\hypersetup{
   colorlinks=true,
   linkcolor=blue,
   citecolor=blue,
   urlcolor=Violet
}

\newcommand{\comm}[1]{}

\def\be{\begin{equation}}
\def\ee{\end{equation}}
\def\ba{\begin{align}}
\def\ea{\end{align}}

\theoremstyle{plain}

\theoremstyle{definition}

\newcommand{\bigast}{\mathop{ \mathlarger{\mathlarger{*}}}}

\begin{document}
\title{Quantum teleportation of quantum causal structures}

\author{Marius Krumm}
\email{marius.krumm@uibk.ac.at}
\affiliation{Institute for Theoretical Physics, University of Innsbruck, Technikerstra{\ss}e 21a, 6020 Innsbruck, Austria}
\affiliation{Institute for Quantum Optics and Quantum Information (IQOQI) Vienna, Austrian Academy of Sciences, Boltzmanngasse 3, 1090 Vienna, Austria}
\affiliation{University of Vienna, Faculty of Physics,  Boltzmanngasse 5, 1090 Vienna, Austria}

\author{Philippe Allard Gu\'{e}rin}
\affiliation{Perimeter Institute for Theoretical Physics, 31 Caroline Street North, Waterloo ON N2L 2Y5, Canada}

\author{Thomas Zauner}
\affiliation{Institute of Science and Technology Austria (ISTA), 3400 Klosterneuburg, Austria}

\author{\v{C}aslav Brukner}
\affiliation{Institute for Quantum Optics and Quantum Information (IQOQI) Vienna, Austrian Academy of Sciences, Boltzmanngasse 3, 1090 Vienna, Austria}
\affiliation{University of Vienna, Faculty of Physics,  Boltzmanngasse 5, 1090 Vienna, Austria}

\begin{abstract}
	Quantum teleportation is a very helpful information-theoretic protocol that allows to transfer an unknown arbitrary quantum state from one location to another without having to transmit the quantum system through the intermediate region. Quantum states, quantum channels, and indefinite causal structures are all examples of quantum causal structures that not only enable advanced quantum information processing functions, but can also model causal structures in nonclassical spacetimes. In this letter, we develop quantum teleportation of arbitrary quantum causal structures, as formalized by the process matrix framework. Instead of teleporting all the physical degrees of freedom that implement the causal structure, the central idea is to just teleport the inputs to and outputs from the operations of agents. The communication of outcomes of Bell state measurements, which is necessary for deterministic quantum teleportation, is not possible for all causal structures that one might wish to investigate. To avoid this problem, we propose partially and fully post-selected teleportation protocols. We prove that our partially post-selected teleportation protocol is compatible with all quantum causal structures, including those that involve indefinite causal order.
\end{abstract}

\date{February 28, 2022}

\maketitle

Quantum teleportation~\cite{Teleportation, TeleportationExperiment1, TeleportationExperiment2} is a protocol from quantum information theory that allows to transfer a quantum state from one location to another, even if the quantum state is unknown and without passing physically the state from one location to another location. Therefore, quantum teleportation is a helpful tool that makes quantum communication and quantum computation more flexible. It allows to exchange the physical system to which a quantum state is assigned and to transfer quantum information without transmitting its physical carrier. Explicit examples for applications are measurement-based quantum computation~\cite{MBQC} and quantum repeaters for long distance quantum communication~\cite{QuantumRepeaters}.

Recently, combining the concepts and tools from quantum information, computer science, and general relativity, reseachers started to study quantum causal structures~\cite{OCB, QuantumCausalModels}. The preparation and distribution of quantum states is an example of such a structure. Quantum circuits are also examples of quantum causal structures, also known in the literature as \emph{quantum combs}~\cite{CombsLong}. Furthermore, quantum theory admits so-called \emph{indefinite causal structures}, exotic causal structures in which the order of quantum gates is affected by quantum uncertainty. Such indefinite causal structures are expected to arise in quantum gravity when the light cone structure is blurred by quantum fluctuations~\cite{Hardy2007} or in the vicinity of large masses prepared in spatial superposition~\cite{ZychThesis, zych2019bell}. A typical example is the \emph{quantum switch}~\cite{Chiribella2013} for which the order of two quantum gates is controlled by a quantum system. It has recently been demonstrated in quantum optics experiments~\cite{rubino2017experimental, Taddei2020, Guo2020, SwitchExpCommunicationComplexity}, and it was shown that its multiparty generalisation (i.e. the n-switch) can reduce sublinearly the computational query complexity~\cite{AraujoSwitchAdvantage, RennerSwitchAdvantage1, RennerSwitchAdvantage2} in certain black box problems and it can exponentially reduce the complexity in quantum communication problems~\cite{SwitchExponentialCommunication}.

This raises the question whether it is possible to teleport quantum causal structures. In the context of quantum computation, this would open up the possibility that a large-scale quantum server implements a large quantum circuit fragment, while clients just need to teleport their quantum states, gates or small-scale circuit fragments, and they do not even have to reveal a classical description of their choices. In the context of quantum gravity, teleportation might help to investigate fragile indefinite causal structures by implementing large parts of the agents' quantum instruments far away.

In this letter, we extend the concept of quantum teleportation to arbitrary quantum causal structures. A naive generalization of quantum teleportation would teleport all the involved physical degrees of freedom used to implement the quantum causal structure. However, using the process matrix framework~\cite{OCB} to formalize quantum causal structure, we show that it is enough to teleport only the inputs and outputs of the quantum causal structure. Since deterministic quantum teleportation requires classical communication of outcomes of Bell state measurements, deterministic teleportation imposes additional causal compatibility conditions that might be incompatible with the quantum causal structure under investigation. To avoid this, we propose partially and fully post-selected teleportation protocols that can be applied to an arbitrary amount of agents. We prove that the partially post-selected teleportation protocols are compatible with all quantum causal structures.

\textbf{Process matrices.} The process matrix formalism~\cite{OCB} is a minimal-assumptions approach that allows to describe quantum causal structures. It models causal interventions as quantum instruments. Let $\mathcal L(\mathcal H)$ be the set of linear operators on a Hilbert space $\mathcal H$. In particular, $\mathcal L(\mathcal H)$ contains the set of density operators over $\mathcal H$. Then a quantum instrument $\{\mathcal M_a\}_{a=1}^n$ is given by a collection of (probabilistic) quantum transformations, i.e. linear, completely-positive, trace-non-increasing maps $\mathcal M_a: \mathcal L(\mathcal H_{\mathrm{in}}) \rightarrow \mathcal L(\mathcal H_{\mathrm{out}})$, such that $\sum_{a=1}^n \mathcal M_a$ is a completely positive and trace-preserving (CPTP) linear map.

It is more convenient to work with the Choi operators~\cite{Choi,Jamiolkovski} associated with the quantum transformations. For an orthonormal basis $\ket{0},\dots,\ket{d_{\mathrm{in}} - 1}$ of the input space, the Choi operator $M$ associated with the linear map $\mathcal M : \mathcal L(\mathcal H_{\mathrm{in}}) \rightarrow \mathcal L(\mathcal H_{\mathrm{out}})$ is defined as $M := (\mathcal I \otimes \mathcal M)(|\mathbb 1 \rangle \rangle \langle \langle \mathbb 1 |)$ where $|\mathbb 1 \rangle \rangle = \sum_{j=0}^{ d_{\mathrm{in}} -1} \ket{j} \otimes \ket{j}$ is an unnormalized maximally entangled state on the input space and a copy, and $\mathcal I$ is the identity channel.

\begin{figure}[hbt]
\includegraphics[width = 0.2\textwidth]{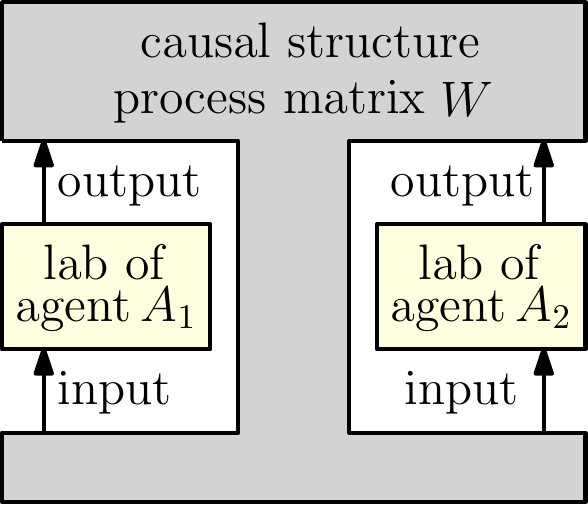}
\caption{The process matrix framework allows to describe quantum causal structures. Causal interventions are modeled as quantum instruments applied by agents in small labs. The causal structure that connects these labs is modeled by the process matrix $W$.}
\label{Figure:W}
\end{figure}

In the process formalism (see Figure \ref{Figure:W}) one imagines several agents $A_1$, \dots, $A_N$ within small labs in which the usual rules of quantum mechanics are valid. Each agent receives a quantum system from the environment, acts on it with their quantum instrument $\mathcal M_{a}^{A_j}$, and then sends it out again. The environment connecting the labs might be incompatible with a definite causal order, e.g. it might be a controlled superposition of causal orders as in the quantum switch~\cite{Chiribella2013, zych2019bell}.

Then, a general process is defined as the most general multi-linear map that maps the instrument choices of the agents to probability distributions (even if the agents share additional entangled states). In the Choi-representation, this can be written as
$p(a_1,\dots, a_N) = \mathrm{Tr}[W^T \cdot (M_{a_1}^{A_1} \otimes \dots \otimes M_{a_N}^{A_N})]$. Here, $a_j$ is the outcome of agent $A_j$ and $W$ is the process matrix that gives the framework its name. Details can be found e.g. in \cite{OCB,WitnessingCausalNonseparability}.




\textbf{General teleportation scheme.} The basic idea of quantum teleportation of causal structures is as follows: Instead of teleporting all the degrees of freedom involved in the physical realization of the causal structure (e.g. a large mass creating a superposition of spacetimes), it is enough to teleport the inputs and outputs of the agents' labs. In this way,  outside parties can obtain access to the correlations produced by a quantum causal structure as if the underlying degrees of freedom were accessible.

\begin{figure}[hbt]
\begin{center}
\includegraphics[width=.5\textwidth]{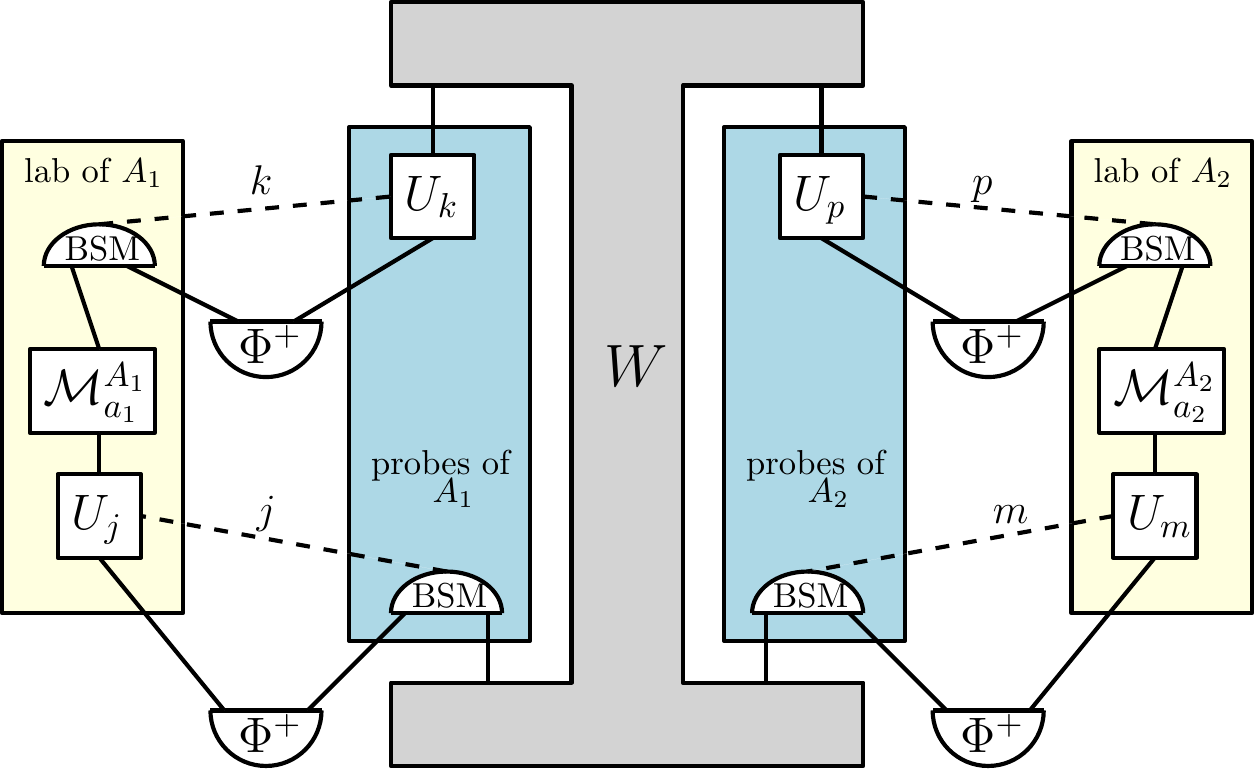}
\caption{Teleportation of a causal structure $W$ for two agents $A_1$ and $A_2$. The protocol sends probes into the causal structure that teleport the inputs to the agents' labs. After the agents have applied their instruments $\mathcal M_{a_j}^{A_j}$ to these inputs, they teleport the outputs back into the causal structure.}
\label{Figure:WTeleportation}
\end{center}
\end{figure}

More specifically, we imagine that the agents and their labs stay outside of the causal structure. Each agent sends probe systems into the causal structure. These probe systems teleport the input from the causal structure to the agents' labs, who then apply their quantum instruments. Afterwards, the agents teleport the outputs of their quantum instruments back into the causal structure. The general teleportation scheme is visualized in Figure \ref{Figure:WTeleportation}.

In detail, the protocol works as follows: Each agent sends two probes into the causal structure. One of these probes has the same dimension as the input from the causal structure (we call it the \emph{input probe}), the other probe has the same dimension as the output to the causal structure (we call it the \emph{output probe}). Furthermore, the corresponding agent has a copy of each probe. Each probe and its copy are in a maximally entangled state $\ket{\Phi^+} := \frac{1}{\sqrt{d}}\sum_j \ket{j}\otimes \ket{j}$, where $d$ is the input or output dimension of the agent's socket in the causal structure. 

Instead of directly applying an arbitrary instrument within the causal structure, only fixed teleportation operations are applied. First, the input from the causal structure is teleported to the lab of the respective agent: A generalized Bell state measurement (BSM), i.e. a measurement in a maximally entangled basis, is applied to the input from the causal structure and the input probe. In the ideal case, the outcome $j$ of the measurement is communicated to the agent on the outside, who applies a correcting unitary $U_j$ to their copy of the probe. 

Then, the agent (here $A_m$) applies their instrument $\mathcal M_{a}^{A_m}$ to the copy of the input probe. Finally, the output of this instrument is teleported from the agent's lab back into the corresponding socket of the causal structure: A generalized Bell state measurement is applied to this output and the copy of the output probe. In the ideal case, the outcome $k$ of this measurement is communicated to the agent's socket of the causal structure, such that a correcting unitary $U_k$ can be applied to the output probe before it is sent as output to the causal structure. For completeness, the detailed protocol of \cite{Teleportation} for teleportation of states in arbitrary dimension (including an example for a maximally entangled basis and corresponding correcting unitaries $U_j$) is given in the Supplementary Material.

\textbf{Post-selection instead of communication.} The full teleportation protocol requires communicating the outcome of the generalized Bell state measurements between the agent's lab and their socket in the causal structure. As we consider several agents, this need for communication with the possibly indefinite causal structure raises questions about the causal and operational situation of the agents and their labs with respect to each other, and with respect to the causal structure. In other words, the use of communication induces additional compatibility conditions for the relation between the agents and the causal structure. 

Before we dive into this issue, we first point that there exist state teleportation protocols (including the one in the Supplementary Material or \cite{Teleportation}) for which there exists one outcome (say $0$) that does not require a correcting unitary, i.e. $U_0 = \mathbb 1$. Furthermore, the protocol can be chosen such that this outcome occurs with probability $\frac{1}{d^2}$, where $d$ is the dimension of the teleported probe. In particular, this probability is independent of the teleported state.

\begin{figure}[hbt]
\includegraphics[width=0.5\textwidth]{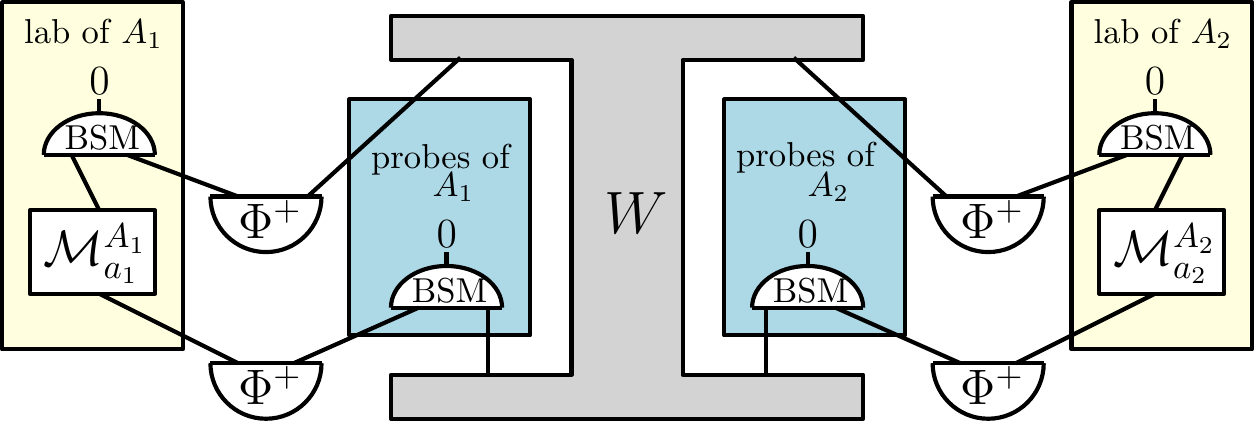}
\caption{The fully post-selected process teleportation protocol. This protocol replaces the outcome communication and correcting  unitaries of Figure \ref{Figure:WTeleportation} with post-selection. Here, the agents' causal relations can be arbitrary and independent of the process $W$.}
\label{Figure:FullyPostSelected}
\end{figure}

This observation creates the opportunity for a post-selection-based protocol that allows the probing of arbitrary causal structures, independently of the agents' own causal situations, see Figure \ref{Figure:FullyPostSelected}.

More specifically, in Figure \ref{Figure:WTeleportation}, instead of communicating the outcomes of the generalized Bell state measurements between the labs and the sockets, the outcomes are just recorded. No correcting unitaries are applied: The agents directly apply their instruments $\mathcal M_{a}^{A_m}$ to the copies of the input probes, and similarly the output probes are sent directly into the causal structure.

At the end, the recorded outcomes and statistics are collected and all statistics for which the outcomes of the generalized Bell state measurements were not $0$ are discarded. For example, one can imagine that the agents are in a quantum causal structure only for the duration of the experiment, and after that they can ``leave'' it (or the structure ceases to exist) and communicate with each other or meet to compare results and determine correlations. The probability for a successful run of this protocol is $\prod_{j=1}^N \frac{1}{d^2_{j,\mathrm{in}} d^2_{j, \mathrm{out} }}$. Here, $N$ is the number of agents, $d_{j,\mathrm{in}}$ the input dimension of agent $j$ and $d_{j, \mathrm{out}}$ the output dimension.

While this post-selected protocol requires a large number of repetitions of the experiment (in particular a large number of copies of the probed causal structure) to be successful with high probability, it massively simplifies the individual runs themselves by dropping the need for communication and relaxing the requirements on the agents' causal situations.

\textbf{Communication and causal structure.} Can we increase the success probability by allowing for outcome communication and correcting unitaries, while imposing stricter causal requirements for the agents?  

The use of communication and correcting unitaries in Figure \ref{Figure:WTeleportation} induces the following causal ordering of the operations of an agent's lab and the operations in the sockets of the causal structure: First, the causal structure provides the input. Then the first generalized Bell state measurement is applied to this input and the input probe. Only then can the agent apply the correcting unitary and the instrument. Then the agent applies the generalized Bell state measurement. Only after this has happened, the correcting unitary in the socket can be applied before the output to the causal structure is provided. 

This description seems to imply that (without post-selection) the causal situation of the agents must be perfectly compatible with the probed causal structure, since some of the operations in the socket happen before the operations of the corresponding agent, while other operations of the socket happen after the agent's operations. In other words, an agent's operations must happen while their socket in the causal structure is ``active''. But in the case of an indefinite causal structure, this means that the agents' labs themselves must be in an indefinite causal relation to each other. 

\textbf{Partial post-selection.} Operationally, in the case of probing an indefinite causal structure, the most appealing scenario would be one in which the agents are outside of the probed indefinite causal structure, i.e. the agents themselves are part of a definite causal structure. We have seen a post-selection protocol that can achieve this goal, but it has a small success probability. However, our arguments from the previous section suggest that this probability cannot be boosted to $1$, in general. However, we can increase the success probability by having a mix of communication and post-selection. The protocol of Figure \ref{Figure:WTeleportation} requires two-way communication between the agent's lab and their socket in the causal structure. This can be relaxed to one-way communication if we post-select exactly one of the generalized Bell state measurements.

\begin{figure}[hbt]
\includegraphics[width = 0.5\textwidth]{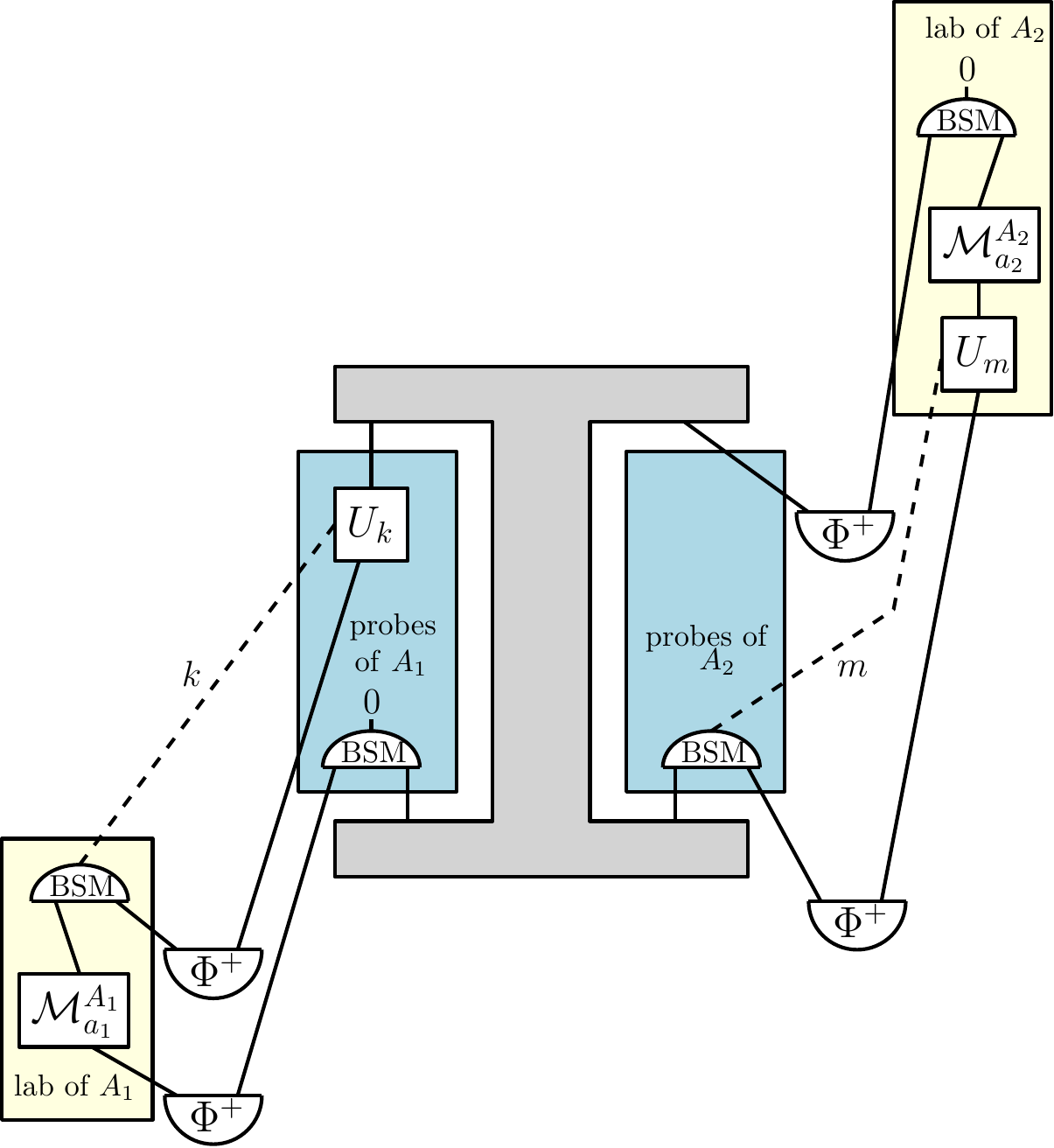}
\caption{Partially post-selected protocols. They only post-select one of the generalized Bell state measurements per party and put the agents into the causal past or future of the process.}
\label{Figure:PartialPostSelected}
\end{figure}

More specifically: Let us say that the first generalized Bell state measurement (i.e. the one in the socket of the causal structure) is post-selected on the outcome $0$ with trivial correction $U_0 = \mathbb 1$. As the corresponding agent does not have to wait for the outcome of the first generalized Bell state measurement, this allows the agent to directly apply their instrument and generalized Bell state measurement in a distant past before the causal structure even arises. The outcome of this second generalized Bell state measurement and the probes must be stored until the agent's socket of the causal structure emerges. The success probability is $\prod_{j=1}^N \frac{1}{d^2_{j,\mathrm{in}} }$ if $N$ agents apply this protocol. This choice is implemented by agent $A_1$ in Figure \ref{Figure:PartialPostSelected}.

Analogously, let us say that the the second generalized Bell state measurement is the post-selected one. Then the outcome of the first generalized Bell state measurement and the copies of the probes can be stored until in a distant future (long after the causal structure existed) the agent applies the correcting unitary, their instrument, and the second generalized Bell state measurement. The post-selection means that the output probe does not have to wait for a correcting unitary, so it can directly be sent as output to the causal structure. The success probability is $\prod_{j=1}^N \frac{1}{d^2_{j, \mathrm{out} }}$ for $N$ agents that follow this protocol. This choice is implemented by agent $A_2$ in Figure \ref{Figure:PartialPostSelected}.

These protocols that mix post-selection and outcome communication in principle allow agents in a definite causal structure to probe an indefinite causal structure. However, there are operational subtleties concerning the communication between the agents and the probed indefinite causal structure: In principle, the signal between the two causal structures could contain information that reveals the causal order of the indefinite causal structure, therefore collapsing the indefinite causal structure. An example of such information could be the arrival time of the communicated system in the socket of the causal structure. In the optical realization of the quantum switch, the photon enters each socket at superimposed times \cite{TimeDelocalized}, either "earlier" or "later", depending on whether the operation in the socket is performed before or after that in the other socket. Now, if the signal sent by the quantum switch to the external agents contains the arrival time of the photon in the sockets, this would cause the collapse of the causal structure to a definite order. On the other hand, if this information is not recorded or erased, the causal structure will remain unaltered.

We point out that the protocol which post-selects all generalized Bell state measurements has no problems related to a collapse, because it does not involve communication between the agents' labs and their sockets in the indefinite causal structure.

In the Supplementary Material we show that the partially post-selected protocols are described by a valid process matrix, which implies that these protocols indeed do give a consistent quantum causal structure.

\begin{figure}[hbt]
	\includegraphics[width = 0.5 \textwidth]{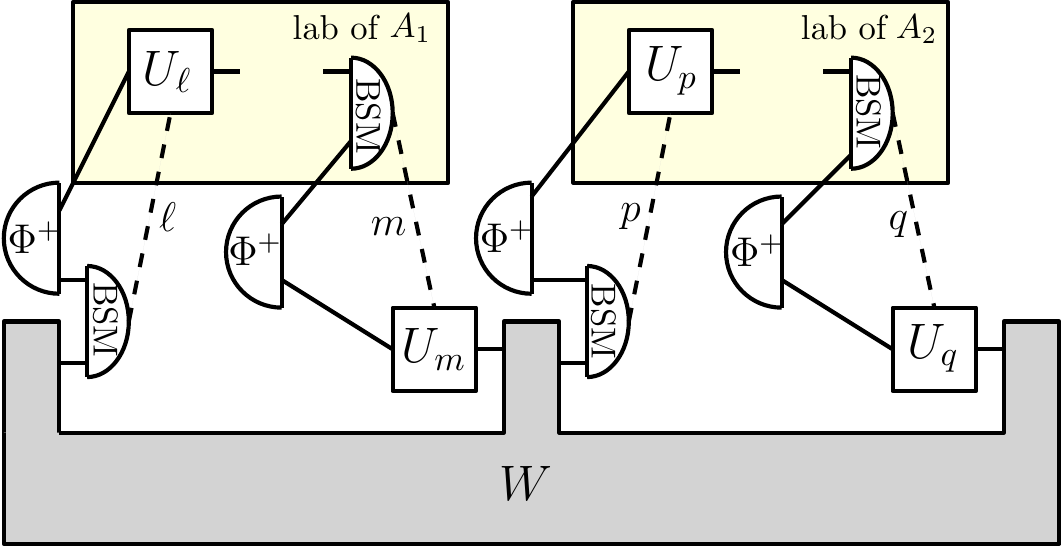}
	\caption{In a definite causal order, i.e. a quantum network or quantum comb~\cite{Chiribella2009,CombsLong}, the deterministic teleportation protocol of Figure \ref{Figure:WTeleportation} can be directly applied by putting the agents in the right order. Alternatively, the post-selected protocols allow to put the agents in arbitrary relation. The agent's instruments are not shown, since definite causal order allows to insert quantum combs into the open sockets of quantum combs.}
	\label{Figure:CombTeleportation}
\end{figure}

\textbf{Teleportation of definite causal structures.} In the context of a definite causal structure or a causal structure with classical control, this issue of collapse induced by outcome communication is of no big concern. First of all, given such a causal structure, it is not difficult to arrange the order of the agents' actions such that they are compatible with this causal structure. In the case of a definite causal order, i.e. a quantum comb or quantum network~\cite{Chiribella2009, CombsLong}, the causal order of the agents just has to agree with the causal order of the comb. If the order of the agents is not fixed in advance, but controlled by classical uncertainty~\cite{WechsQuantumControlledOrder}, then it is enough to send a classical signal to the agents when it is their turn to act. The classical uncertainty can be restored by forgetting the order of the agents in each run, or averaging. Therefore, the deterministic teleportation protocol from Figure \ref{Figure:WTeleportation} can be directly applied to such causal structures by simply putting the agents in the right order, see Figure \ref{Figure:CombTeleportation}.

However, our partial or full post-selection protocols from the previous sections also enable the probing of definite causal structures or causal structures with classical control if the agents have a definite causal order that is not compatible with the causal structure under investigation: In those cases, we do not even have to worry about a possible collapse of indefiniteness caused by outcome communication.

\textbf{Conclusions.} In this letter, we have extended the concept of quantum teleportation to arbitrary causal structures. We presented an operational scenario that sends probes into the causal structure and uses these probes to teleport the agents' quantum instruments to the investigated causal structure. We observed that the need for communication of Bell state measurement outcomes leads to additional causal consistency requirements between the agents and the probed causal structure. To avoid these causal consistency requirements, we introduced modified teleportation protocols that post-select some or all Bell state measurements on an outcome which does not require a correcting unitary. We pointed out potential subtleties concerning a collapse of indefinite causal structures induced by outcome communication. 

Our results may be useful in situations where agents do not have direct access to (in)definite causal structures, but wish to use these resources to enhance quantum computation or quantum communication. They can also be used to test indefinite causal structures in future quantum gravity scenarios.

\begin{acknowledgments}
\textbf{Acknowledgements.} This letter contains and generalizes results that also appear in the Master thesis of one of the authors of this letter (Thomas Zauner)~\cite{ZaunerMaster}. We acknowledge financial support from the Austrian Science Fund (FWF) through BeyondC (F7103-N38), the project no. I-2906, as well as support by the John Templeton Foundation through grant 61466, The Quantum Information Structure of Spacetime (qiss.fr), the Foundational Questions Institute (FQXi) and the research platform TURIS. The opinions expressed in this publication are those of the authors and do not necessarily reflect the views of the John Templeton Foundation. Research at Perimeter Institute is supported in part by the Government of Canada through the Department of Innovation, Science and Economic Development Canada and by the Province of Ontario through the Ministry of Colleges and Universities.
\end{acknowledgments}




\onecolumngrid

\appendix
\section*{Supplementary material}
\section{Review: Quantum teleportation of states in higher-dimensional Hilbert spaces}
\label{Appendix:HigherDimension}

As many lectures and textbooks only describe the quantum teleportation protocol of states for two-dimensional systems, for completeness we will explain the quantum teleportation protocol for arbitrary dimension. While there are many teleportation schemes~\cite{WernerTeleportation}, we will explain the specific scheme of Reference \cite{Teleportation}.

Bennett et al.~\cite{Teleportation} generalize the usual Bell basis to arbitrary dimension $d$ by introducing an orthonormal basis consisting of maximally entangled states $\ket{\psi_{nm}}$, $0\le n,m \le d-1$ given by
\begin{align}
	\ket{\psi_{nm}} = \frac{1}{\sqrt{d}}\sum_{j=0}^{d-1} e^{2\pi i j n / d} \ket{j}\otimes \ket{(j+m) \text{ mod } d} \label{Equation:BellStates}
\end{align}
Indeed, this is a set of $d^2$ orthonormal states as one can directly verify (leaving $\text{mod } d$ implicit):
\begin{align*}
	\braket{\psi_{n' m'} | \psi_{n m}} = \frac{1}{d} \sum_{j,k=0}^{d-1} e^{2\pi i (j n - k n') / d} \braket{k | j} \braket{k+m' | j + m} = \frac{1}{d} \sum_{k=0}^{d-1} e^{2\pi i k (n - n') / d} \  \delta_{m,m'} = \delta_{n,n'} \delta_{m,m'}
\end{align*}
As generalization of the Pauli matrices, Bennett et al.~\cite{Teleportation} introduce the unitary operators 
\begin{align*}
	U_{nm} := \sum_{k=0}^{d-1} e^{2 \pi i k n / d} \ket{k}\bra{(k+m) \text{ mod } d}.
\end{align*}
Now let us consider three $d$ dimensional Hilbert spaces $A, A', B$, with $A,A'$ in possession of agent Alice and $B$ in possession of Bob. Furthermore, on $A'B$ we have the maximally entangled state $\ket{\psi_{00}}= \frac{1}{\sqrt{d}} \sum_k \ket{k}_{A'} \ket{k}_B$. There is an arbitrary state $\ket{\phi}_A$ that is supposed to be teleported from Alice to Bob.

Now, Alice performs a generalized Bell basis measurement in the basis $\ket{\psi_{nm}}_{AA'}$ on her systems $A$ and $A'$. This is described by the quantum instrument $\{\mathcal{B}_{nm}\}_{n,m=0}^{d-1} $ with 
\begin{align*}
	\mathcal B_{nm}: \ \rho_{A A' B} \mapsto (\ket{\psi_{nm}}\bra{\psi_{nm}}_{AA'}\otimes  \mathbb 1_{B})\cdot \rho \cdot (\ket{\psi_{nm}}\bra{\psi_{nm}}_{AA'}\otimes  \mathbb 1_{B})
\end{align*} 
In particular, upon receiving outcome $n,m$ Alice will see the post-measurement state $\ket{\psi_{nm}}_{AA'}$ and it factorizes from Bob's post-measurement state. Since we consider states of the form $\rho = \ket{\phi}\bra{\phi}_{A} \otimes \ket{\psi_{00}}\bra{\psi_{00}}_{A'B}$, Bob's post-measurement state is thus given by (up to normalization) $\bra{\psi_{nm}}_{AA'} (\ket{\phi}_A \otimes \ket{\psi_{00}}_{A'B})$. Expanding $\ket{\phi}_A = \sum_k \phi_k \ket{k}_A$ we thus find:
\begin{align*}
	\bra{\psi_{nm}}_{AA'} (\ket{\phi}_A \otimes \ket{\psi_{00}}_{A'B}) = \frac{1}{d}\sum_{j} e^{- 2\pi i j n / d} \sum_{k, p} \phi_p \braket{j | p} \braket{j+m | k} \ket{k}_B = \frac{1}{d} \sum_j e^{- 2 \pi i j n / d} \phi_j \ket{ j +m }_B
\end{align*}
Now, Alice tells Bob the outcome $(n,m)$ and Bob applies the unitary $U_{nm}$ to obtain the state $\frac{1}{d} \sum_j \phi_j \ket{j}_B = \frac{1}{d} \ket{\phi}_B$. The prefactor $\frac{1}{d}$ expresses the fact that the outcome $(n,m)$ in the generalized Bell basis measurement only occurs with probability $\frac{1}{d^2}$.

\section{Proof that the partially post-selected process teleportation protocols correspond to a valid process matrix}
Even though the teleportation protocols reproduce the statistics of the original process, one has to check that the teleportation scenario is embedded into a valid (quantum) causal structure. In principle, the agents could choose to not follow the teleportation procedure and apply any other quantum instrument instead. Nonetheless, the input-output statistics should lead to well-defined probabilities. To ensure that this is the case, we prove that the extended process matrix including the probes and the outside agents is itself also a valid process.

For the fully post-selected protocol, there is no actual communication between the outside agents and the sockets of the original process. Therefore, adding the outside agents still leads to a valid process matrix. Furthermore, process matrices are explicitly defined to allow agents to access additional entangled quantum states, such as our probes and their copies. Therefore, adding the probes and their copies also leads to a valid process matrix.

The situation is more complicated for the partially post-selected protocols because of the additional communication involved. Therefore it is crucial to check that, nonetheless, the complete scenario is described by a valid process matrix.

First of all, we can ignore the initial distribution of the probes (and their copies). Just as explained before, if we can show that the addition of the outside agents and their communication leads to a valid process, we can freely add additional entangled systems (such as the initial probes) for all agents to act on and still get a valid process.

We introduce the index sets $P$ and $F$. Here, $j \in P$ means that agent $j$ is in the causal past of $W$, while $j \in F$ means that agent $j$ is in the causal future of $W$.

We model the outcome information exchanged between the outside agents and the probes on the inside as being encoded into a quantum system and being transmitted via an identity channel, with Choi operator $\ket{\mathbb 1}\bra{\mathbb 1}$. If outside agent $j$ is in the causal past of $W$, we extend the input of agent $j$ of the original process $W$ (i.e. the inner agent) to have an additional input space $I_j^M$ for receiving the message, while the output space $\tilde{O}_j$ of outside agent $j$ is given solely by the message. If outside agent $j$ is in the causal future of $W$, we give agent $j$ of the original process $W$ (i.e. the inner agent) an additional output space $O_j^M$ for the message, and the input space $\tilde{I}_j$ of outside agent $j$ is given completely by the message (remember that we can ignore the distribution of the probes).

\begin{figure}[h!]
\includegraphics[width= 0.6 \textwidth]{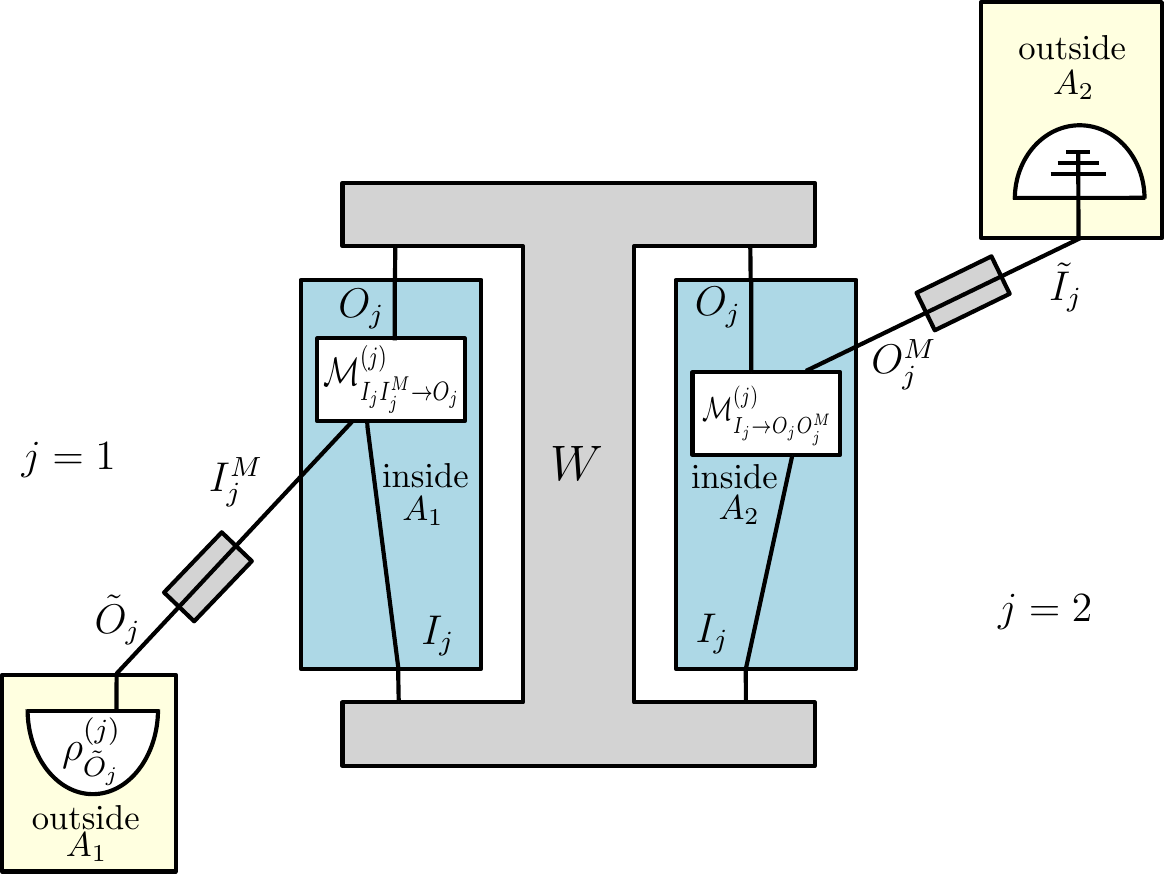}
\caption{\small The graph that must be shown to be a valid process matrix to prove that partially post-selected teleportation (compare Figure \ref{Figure:PartialPostSelected}) corresponds to a valid quantum causal structure. Non-negativity of probabilities is clear because all involved operators are positive. Only normalization of probabilities must be shown. For this purpose, the initial distribution of entangled probes can be ignored. Thus, it is enough to show that the causal structure in the figure corresponds to a valid process matrix that gives probabilities with normalization $1$. The outcome communication for correcting the teleportation happens via quantum systems $I_j^M$ and $O_j^M$. The deterministic inputs are quantum states $\rho_{\tilde{O}_j}^{(j)}$, the only deterministic POVM element is the trace. Also shown are channels $\mathcal M^{(j)}_{I_j I_j^M \rightarrow O_j}$ and $\mathcal M^{(j)}_{I_j \rightarrow O_j O_j^M}$ of the inside agents. Grey boxes belong to the extended process matrix, while white, yellow and blue boxes belong to the agents. Identity channels are represented as boxes with a wire through them.}
\label{Figure:ValidW}
\end{figure}

With this notation, we have to prove the validity of the following process, see Figure \ref{Figure:ValidW}:
\begin{align}
	V:= W \bigotimes_{j \in P} \ket{\mathbb 1}\bra{\mathbb 1}_{\tilde{O}_j \rightarrow I_j^M}  \bigotimes_{j \in F} \ket{\mathbb 1}\bra{\mathbb 1}_{O^M_j \rightarrow \tilde{I}_j} 
\end{align}

First of all, we notice that the extended process $V$ is a positive operator, and therefore will lead to non-negative probabilities on all quantum instruments. This holds true even if we add additional entangled systems for all agents to act on, like the probes, because the extended process matrix is still a positive operator.

Therefore, we only have to prove that $V$ gives probabilities that are normalized to $1$ for all choices of instruments. As expressed in Appendix B of \cite{WitnessingCausalNonseparability} (see Footnote 15) for bipartite process matrices, we can ignore the possibility of additional entangled systems. For completeness, in the next section of the Supplementary Material we will prove that this statement extends to an arbitrary amount of agents.

As we are checking for normalization, we only need to consider deterministic agent operations, i.e. quantum channels. The outside agents $j \in F$ in the causal future of $W$ have no output and therefore have only one deterministic channel given by $\rho_{\tilde{I}_j} \mapsto \mathrm{Tr}[\rho_{\tilde{I}_j}]$. The Choi operator of this map is $\mathbb 1_{\tilde{I}_j}$. The outside agents $j \in P$ in the causal past of $W$ have no input and therefore their only channel is to output a normalized state $\rho^{(j)}_{\tilde{O}_j}$.

Therefore we have to prove that 
\begin{align}
	1 \stackrel{?}{=} \mathrm{Tr}\left[  \Big(W  \bigotimes_{j \in P} \ket{\mathbb 1}\bra{\mathbb 1}_{\tilde{O}_j \rightarrow I_j^M}  \bigotimes_{j \in F} \ket{\mathbb 1}\bra{\mathbb 1}_{O^M_j \rightarrow \tilde{I}_j} \Big)\cdot \Big( \bigotimes_{j \in P} \rho^{(j)}_{\tilde{O}_j}  \bigotimes_{j \in F} \mathbb 1_{\tilde{I}_j}  \bigotimes_{j \in P} M^{(j)}_{I_j I_j^M \rightarrow O_j}  \bigotimes_{j \in F} M^{(j)}_{I_j  \rightarrow O_j O_j^M} \Big)^T\right] \label{Equation:IsNormalized}
\end{align}
Here, $M^{(j)}_{I_j I_j^M\rightarrow O_j}$ and $M^{(j)}_{I_j \rightarrow O_j O_j^M}$ are the Choi operators of arbitrary (deterministic) quantum channels of the inside agents. 

To avoid a convoluted notation, we introduce the link product of \cite{CombsLong, Chiribella2009}. It is defined as:
\begin{align}
	M_{AB} * N_{BC} := \mathrm{Tr}_B [M_{AB}^{T_{B}} \cdot N_{BC}]
\end{align}

Ref. \cite{CombsLong, Chiribella2009} has shown that it is the correct way to link together Choi operators of maps, or circuit fragments. Furthermore, it was shown that it is commutative and associative. With it, Equation \eqref{Equation:IsNormalized} reads:
\begin{align}
	1 \stackrel{?}{=}  W  \bigast_{j \in P}  \ket{\mathbb 1}\bra{\mathbb 1}_{\tilde{O}_j \rightarrow I_j^M}  \bigast_{j \in F}  \ket{\mathbb 1}\bra{\mathbb 1}_{O^M_j \rightarrow \tilde{I}_j}   \bigast_{j \in P} \rho^{(j)}_{\tilde{O}_j} \bigast_{j \in F} \mathbb 1_{\tilde{I}_j} \bigast_{j \in P} M^{(j)}_{I_j I_j^M \rightarrow O_j} \bigast_{j \in F} M^{(j)}_{I_j  \rightarrow O_j O_j^M}  \label{Equation:IsNormalizedLinking}
\end{align}
Using commutativity and the fact that the link product is the right way to combine circuit fragments, we can contract the identity channels to find:
\begin{align}
	1 \stackrel{?}{=}  W   \bigast_{j \in P} \Big( M^{(j)}_{I_j I_j^M \rightarrow O_j} * \rho^{(j)}_{I_j^M} \Big)  \bigast_{j \in F} \Big(\mathbb 1_{O_j^M} * M^{(j)}_{I_j  \rightarrow O_j O_j^M}  \Big) \label{Equation:IsNormalizedContracted}
\end{align}
Here, $M^{(j)}_{I_j I_j^M \rightarrow O_j} * \rho^{(j)}_{I_j^M} $ is the Choi operator of the CPTP map that arises by fixing the input of $M^{(j)}_{I_j I_j^M \rightarrow O_j}$ on $I_j^M$ to be $\rho^{(j)}_{I_j^M} $. Similarly, $\mathbb 1_{O_j^M} * M^{(j)}_{I_j  \rightarrow O_j O_j^M}$ is the Choi operator of the CPTP map that is obtained by tracing out the $O_j^M$-output of $M^{(j)}_{I_j  \rightarrow O_j O_j^M}$. If we define these new CPTP maps to have Choi operators $N_{I_j \rightarrow O_j}$ and $Q_{I_j \rightarrow O_j}$, we find that  Equation~\eqref{Equation:IsNormalized} is equivalent to 
\begin{align}
	1 = \mathrm{Tr}\Big[W^T \cdot \Big(\bigotimes_{j \in P} N_{I_j \rightarrow O_j}  \bigotimes_{j\in F} Q_{I_j \rightarrow O_j} \Big) \Big]
\end{align}
which is always true, because $W$ is a valid process matrix.

\section{Normalization of process probabilities can be checked without introducing additional shared states}
In this section, we prove that the proper normalization of the process matrices can be verified without introducing additional states shared by the agents. Ref. \cite{WitnessingCausalNonseparability} explicitly stated this for two parties, and here we prove this claim for an arbitrary amount of parties.

Consider agents labeled $j$ with input spaces $I_j$ and output spaces $O_j$. We collectively label the input spaces as $I$ and the output spaces as $O$. A process matrix $W^{IO}$ is a positive operator on all of the $I_j$ and $O_j$ such that
\begin{align}
	\mathrm{Tr}\left[ \left( \bigotimes_j M_j^{I_j \tilde{I}_j O_j} \right) \cdot \left( W^{IO} \otimes \rho^{\tilde{I}} \right)\right] = 1 \label{Equation:NormalizationWithAncillas}
\end{align}  
for all CPTP maps $\mathcal M^{I_j \tilde{I}_j O_j}_j : I_j \tilde{I}_j \rightarrow O_j$ with corresponding Choi operators $M^{I_j \tilde{I}_j O_j}_j$, and for all density matrices $\rho^{\tilde{I}}$ on arbitrary ancillary input spaces $\tilde{I}_j$ collectively labeled as $\tilde{I}$.

Here, positivity of $W^{IO}$ already guarantees that probabilities are non-negative, even if additional $\rho^{\tilde{I}}$ are considered. Therefore, one only has to check normalization of probabilities. For this purpose, it is enough to consider deterministic instruments, i.e. CPTP maps.

However, so far it seems that that we have to consider (Choi operators of) CPTP maps $M^{I_j \tilde{I}_j O_j}_j$ on $I_j$ and all possible ancillas $\tilde{I}_j$. In this section, we prove that we can choose the $\tilde{I}_j$ to be trivial, i.e. that we do not need to add $\rho^{\tilde{I}}$. In other words, we show that if 
\begin{align}
	\mathrm{Tr}\left[ \left( \bigotimes_j M_j^{I_j O_j} \right) \cdot  W^{IO} \right] = 1 \label{Equation:NormalizationWithoutAncillas}
\end{align}  
holds for all CPTP maps $\mathcal M^{I_j  O_j}_j : I_j  \rightarrow O_j$ with corresponding Choi operators $M^{I_j O_j}_j$, then also normalization as in Equation \eqref{Equation:NormalizationWithAncillas} holds for all $\tilde{I}_j$, all density matrices $\rho^{\tilde{I}}$ and all CPTP maps $\mathcal M^{I_j \tilde{I}_j O_j}_j : I_j \tilde{I}_j \rightarrow O_j$ with corresponding Choi operators $M^{I_j \tilde{I}_j O_j}_j$.

So, let us assume that $W^{IO}$ already satisfies the weaker statement as in Equation \eqref{Equation:NormalizationWithoutAncillas}. We will prove that 
\begin{align}
	\mathrm{Tr}_{IO}\Big[ (W^{IO} \otimes \mathbb 1^{\tilde{I}} ) \cdot \bigotimes_j M_j^{I_j \tilde{I}_j O_j} \Big] = \mathbb 1^{\tilde I }  \label{Equation:PartialNormalization}
\end{align} 
for all Choi operators $M_j^{I_j \tilde{I}_j O_j}$ of CPTP maps $I_j \tilde{I}_j \rightarrow O_j$. This implies that 
$\mathrm{Tr}\Big[ (W^{IO} \otimes \rho^{\tilde{I}} ) \cdot \bigotimes_j M_j^{I_j \tilde{I}_j O_j} \Big] = \mathrm{Tr} \rho^{\tilde I }  = 1$, as demanded by Equation \eqref{Equation:NormalizationWithAncillas}. Here, we made use of the identity $\mathrm{Tr}_A[ (Q_A \otimes R_B)\cdot H_{AB}] = R_B \cdot \mathrm{Tr}_A[ (Q_A \otimes \mathbb 1_B)\cdot H_{AB}]$. 
Now, we decompose $M_j^{I_j \tilde{I}_j O_j}$ into a normalization term and the rest. We introduce the notation $_A H := \frac{\mathbb 1_A}{d_A} \otimes \mathrm{Tr}_A H$ and $_{[1-A]} H := H - _A H$. Since $M_j^{I_j \tilde{I}_j O_j}$ is the Choi operator of a CPTP map, it satisfies $_{O_j} M_j^{I_j \tilde{I}_j O_j}= \frac{\mathbb 1^{O_j}}{d_{O_j}} \otimes \mathbb 1^{I_j, \tilde{I}_j} = \frac{\mathbb{1}}{d_{O_j}}$. Therefore, we can write $M_j^{I_j \tilde{I}_j O_j} = \frac{\mathbb{1}}{d_{O_j}} + _{[1-O_j]} M_j^{I_j \tilde{I}_j O_j}$. The other way round, for arbitrary Hermitian $N_j$ on $I_j \tilde{I_j} O_j$, we find that $K_j:=\frac{\mathbb 1}{d_{O_j}} + _{[1-O_j]} N_j$ satisfies the normalization condition of CPTP maps, i.e. $\mathrm{Tr}_{O_j} K_j = \mathbb 1_{I_j, \tilde{I}_j}$.

Therefore, we see that Equation \eqref{Equation:PartialNormalization} is equivalent to 
\begin{align}
	\mathrm{Tr}_{IO}\Big[ \bigotimes_j \Big(\frac{\mathbb{1}^{I_j \tilde{I}_j O_j}}{d_{O_j}} + _{[1-O_j]} N_j\Big) \cdot (W^{IO} \otimes \mathbb 1^{\tilde{I}})\Big] = \mathbb 1^{\tilde{I} } \label{Equation:NoDistributiveLawYet}
\end{align}
for all Hermitian operators $N_j$ on $I_j \tilde{I}_j O_j$. We introduce two index sets $\mathcal I$ and $\mathcal R$. Then, using the distributive law, we sum over all bipartitions $\mathcal I \cup \mathcal R = \{ 1,2,\dots N\}$, with $N$ the number of agents:

\begin{align}
	\sum_{\mathcal I, \mathcal R \ | \ \mathcal I \cup \mathcal R = \{ 1,2,\dots N\} } \mathrm{Tr}_{IO}\Bigg[  \left(\bigotimes_{j \in \mathcal I} \frac{\mathbb{1}_{I_j \tilde{I}_j O_j}}{d_{O_j}} \bigotimes_{j \in \mathcal R} {} _{[1-O_j]} N_j \right) \cdot (W^{IO} \otimes \mathbb 1^{\tilde{I}})\Bigg] = \mathbb 1^{\tilde{I} } \label{Equation:DistributiveNormalization}
\end{align}

Equation \eqref{Equation:DistributiveNormalization} is equivalent to Equation \eqref{Equation:NoDistributiveLawYet}. We will now make multiple use of the identity $\mathrm{Tr}_A[ (Q_A \otimes R_B)\cdot H_{AB}] = R_B \cdot \mathrm{Tr}_A[ (Q_A \otimes \mathbb 1_B)\cdot H_{AB}]$. First, we use it to derive another helpful identity:

\begin{align}
\mathrm{Tr}_{AB}[ _{[1-A]} H \cdot Q] =& \mathrm{Tr}_{AB}[ (H- \frac{\mathbb 1_A}{d_A} \otimes \mathrm{Tr}_A H) \cdot Q] = \mathrm{Tr}_{AB}[ H \cdot Q]  - \mathrm{Tr}_{AB}[ ( \frac{\mathbb 1_A}{d_A} \otimes \mathrm{Tr}_A H) \cdot Q] \nonumber \\ =& \mathrm{Tr}_{AB}[ H \cdot Q]  - \frac{1}{d_A}\mathrm{Tr}_{B}[ ( \mathrm{Tr}_A H) \cdot \mathrm{Tr}_A Q]  = \mathrm{Tr}_{AB}[ H \cdot Q] -\mathrm{Tr}_{AB}[ H \cdot (\frac{\mathbb 1_A}{d_A} \otimes \mathrm{Tr}_A Q)] \nonumber \\
=& \mathrm{Tr}_{AB}[ H \cdot _{[1-A]} Q] & 
\end{align}

With this insight, we can rewrite Equation \eqref{Equation:DistributiveNormalization} as

\begin{align}
	\sum_{\mathcal I, \mathcal R \ | \ \mathcal I \cup \mathcal R = \{ 1,2,\dots N\} } \mathrm{Tr}_{IO}\Bigg[  \left(\bigotimes_{j \in \mathcal I} \frac{\mathbb{1}_{I_j \tilde{I}_j O_j}}{d_{O_j}} \bigotimes_{j \in \mathcal R} N_j \right) \cdot \big( {} _{\prod_{j \in \mathcal R} [1-O_j]} W^{IO} \otimes \mathbb 1^{\tilde{I}} \big) \Bigg] = \mathbb 1^{\tilde{I} } 
\end{align}

and evaluating the partial traces for $\mathcal I$ gives us:

\begin{align}
	\sum_{\mathcal I, \mathcal R \ | \ \mathcal I \cup \mathcal R = \{ 1,2,\dots N\} } \prod_{m \in \mathcal I } \frac{1}{d_{O_m}}\cdot \mathrm{Tr}_{I_k O_k | k \in \mathcal R}\Bigg[   \bigotimes_{j \in \mathcal R} N_j \cdot \big( {} _{\prod_{j \in \mathcal R} [1-O_j]} \mathrm{Tr}_{O_p I_p | p \in \mathcal I} W^{IO} \otimes \mathbb 1^{\tilde{I}} \big) \Bigg] = \mathbb 1^{\tilde{I} } 
\end{align}

Ref. \cite{WitnessingCausalNonseparability} has shown that $_{\prod_{j \in \mathcal R} [1- O_j] \cdot \prod_{k \in \mathcal I} I_k O_k} W^{IO} = 0$ for all non-empty $\mathcal R$. Therefore, only the summand with $\mathcal R = \emptyset $ remains and we find

\begin{align}
	   \prod_{j} \frac{1}{d_{O_j}}\cdot \mathrm{Tr}_{IO}[W^{IO}] \otimes \mathbb 1^{\tilde{I}}  = \mathbb 1^{\tilde{I} } 
\end{align}

Since $\mathrm{Tr}_{IO}[W^{IO}] = \prod_j d_{O_j}$, this last equation is satisfied for all valid process matrices, and so is the equivalent Equation \eqref{Equation:NormalizationWithAncillas}.

\section{Formalization of the partially post-selected process teleportation protocol and proof that it gives the right statistics}
In the main text, we provided an operational description of the partially post-selected process teleportation protocol. In this section, we will provide a mathematical formalization of the protocol in the process matrix formalism, using the Choi isomorphism and the link product. After providing this formalization, we will prove that the proposed partially post-selected process teleportation protocol achieves its goal of teleporting the agents' instruments into the original process. This is physically remarkable, because some of the agents involved in the teleportation protocol may be part of an indefinite causal structure, while others are outside. Despite the physically exotic situation of some of the agents, we will see that the calculations reduce to those of usual quantum teleportation as in the first section of the Supplementary Material.

The basic strategy is to identify the mathematical occurrence of the link product and then successively apply known rules of the link product such that we can replace the full extended scenario of Figure  \ref{Figure:PartialPostSelected} with the original causal structure given by the process matrix $W$, such that the agents only apply the teleportation operations within their own labs. Then the first section of the Supplementary Material tells us that indeed the teleportation protocol reproduces the statistics as if teleportation was not applied (i.e. as if the outside agents directly inserted their main instrument into the original process matrix $W$). 

\begin{figure}[h!]
	\includegraphics[width= 0.6 \textwidth]{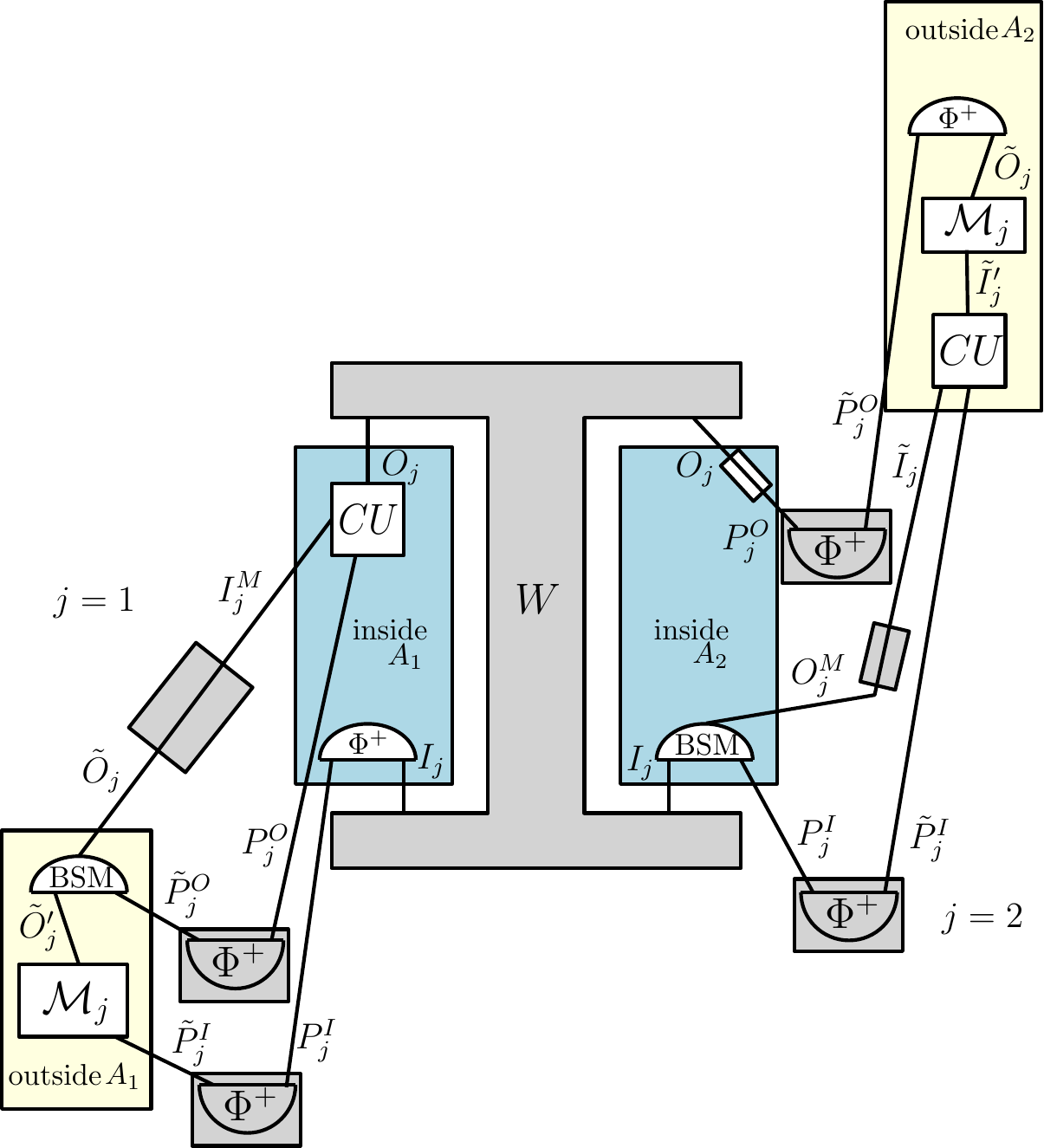}
	\caption{\small This figure shows the Hilbert space labels we use in the partially post-selected protocol of Figure \ref{Figure:PartialPostSelected}. Grey boxes are part of $W_{ext}$, white, yellow and blue boxes are implemented by agents. Identity channels are represented just as boxes with a wire through them. The Bell state measurement $\mathrm{BSM}$ now sends the outcome as a message with a quantum system. $CU$ is an instrument that measures the (quantum) message to decide which correcting unitary $U_m$ to apply to make a teleportation deterministic.}
	\label{Figure:HilbertSpaces}
\end{figure}

First, some notation, compare to Figure \ref{Figure:HilbertSpaces}: The original agents of the process $W$ are labeled by $j \in \{1,2,\dots N\}$, and their input and output spaces are $I_j$ and $O_j$.
Again, we consider an index set $P$ for external agents that are in the causal past of $W$, and an index set $F$ for external agents in the causal future of $W$. For each agent $j$, there are four probes. Two of them are described by Hilbert spaces $P_j^I$ and $\tilde{P}_j^I$ which are copies of $I_j$. These probes are prepared in the normalized maximally entangled state $|\phi^+ \rangle_{P_j^I \tilde{P}_j^I} := \frac{1}{\sqrt{d_{j,\mathrm{in}}}} \sum_{k=0}^{d_{j,\mathrm{in}} - 1} \ket{k}_{P_j^I} \ket{k}_{\tilde{P}_j^I}$, and $d_{j, \mathrm{in}}$ the dimension of $I_j$. Similarly, for each agent $j$ there are two probes with Hilbert spaces $P_j^O$ and $\tilde{P}_j^O$ that are copies of $O_j$, and the probes are prepared in the maximally entangled state $|\phi^+ \rangle_{P_j^O \tilde{P}_j^O} := \frac{1}{\sqrt{d_{j,\mathrm{out}}}} \sum_{k=0}^{d_{j,\mathrm{out}} - 1} \ket{k}_{P_j^O} \ket{k}_{\tilde{P}_j^O}$. $\ket{\phi^+}$ corresponds to the Bell state $\ket{\psi_{00}}$ from the first section of the Supplementary Material. In the cases $j \in P$ and $j \in F$, we introduce Hilbert spaces $\tilde{O}'_j$ and $\tilde{I}'_j$, respectively, to label a direct output/input of $\mathcal M_j$. Here, $\mathcal M_j$ is the quantum transformation that agent $j$ actually wants to implement in $W$. 

The probes labeled $\tilde{P}_j^{I,O}$ are given to the outside agents, the probes labeled $P_j^{I,O}$ are given to the inside agents (in addition to the input $I_j$ provided by $W$). Furthermore, for the agents $j \in P$, there is an identity channel for the (quantum) message from the output space $\tilde{O}_j$ of the outside agent $j$ to an additional input space $I_j^M$ of the inside agent. Meanwhile, for the agents $j \in F$, there is an identity channel for the (quantum) message from an additional output space $O_j^M$ of the inside agent $j$ to the input space $\tilde{I}_j$ of the outside agent.

The initial probes are prepared independently of the process, and the message identity channels connect labs, but not the process. Therefore, the extended process matrix for this scenario is then (as can also be seen by the fact that the link product of \cite{CombsLong} reduces to the tensor product for non-overlapping Hilbert spaces):
\begin{align}
	W_{ext} := W \bigotimes_{j=1}^N \Big(\ket{\phi^+}\bra{\phi^+}_{P_j^I \tilde{P}_j^I} \otimes \ket{\phi^+}\bra{\phi^+}_{P_j^O \tilde{P}_j^O} \Big) \bigotimes_{j \in P} \ket{\mathbb 1} \bra{\mathbb 1}_{\tilde{O}_j I_j^M}  \bigotimes_{j \in F} \ket{\mathbb 1}\bra{\mathbb 1}_{O^M_j \tilde{I}_j}
\end{align}

Now, let us turn to the agents. The inner agents apply a Bell state measurement in the basis of Eq. \eqref{Equation:BellStates} to their input at $I_j$ and the input probe $P_j^I$. In the case $j \in P$, this measurement is post-selected on the outcome $0$ which just applies the POVM element $\ket{\phi^+}\bra{\phi^+}_{I_j P_j^I}$. In the case $j \in F$, the inner agent gets a result $m_j$ (which represents a double index in notation of Eq. \eqref{Equation:BellStates}). This can be represented by an instrument applying Kraus operator $\mathcal{BSM}_{m_j} := \ket{m_j}_{O_j^M} \bra{\psi_{m_j}}_{I_j P_j^I}$ upon outcome $m_j$. Its Choi operator is given by $(BSM_{m_j})_{O_j^M I_j P_j^I} := \ket{m_j}\bra{m_j}_{O_j^M}\otimes \ket{\psi^*_{m_j}}\bra{\psi^*_{m_j}}_{I_j P_j^I}$ with $^*$ complex conjugation in the computational basis. 

Now we consider the outside agents. In the case $j \in P$, the outside agent directly applies their quantum transformation $\mathcal M_j: \tilde{P}_j^I \rightarrow \tilde{O}'_j$ to the input system $\tilde{P}_j^I$, represented by Choi operator $(M_j)_{\tilde{P}_j^I \tilde{O}'_j}$.  Then the outside agent applies the Bell state measurement with outcome $m_j$ represented by $(BSM_{m_j})_{\tilde{O}_j \tilde{O}'_j \tilde{P}^O_j}$. In the case $j \in F$, the outside agent measures the message received on $\tilde{I}_j$ and applies the correcting unitary $U_{m_j}$. This can be modeled via an instrument that applies the Kraus operator $(U_{m_j} )_{\tilde{P}_j^I  \rightarrow \tilde{I}'_j} \otimes\bra{m_j}_{\tilde{I}_j}$. We call its Choi operator $(CU_{m_j})_{\tilde{I}_j \tilde{P}_j^I \tilde{I}'_j }$. Then, the agent applies their instrument $\mathcal M_j : \tilde{I}'_j \rightarrow \tilde{O}_j$. Finally, the outside agent post-selects the Bell state POVM element $\ket{\phi^+}\bra{\phi^+}_{\tilde{O}_j \tilde{P}_j^O}$.

Now, we turn to the inside agents again. In the case $j \in P$, the inside agent applies the unitary correction dependent on the received message, represented by $(CU_{m_j})_{I_j^M P_j^O  O_j}$. In the case $j \in F$, the inside agent just applies the identity channel $\ket{\mathbb 1}\bra{\mathbb 1}_{P_j^O O_j}$.

Now we collect these steps. Within an agents lab, everything has a definite causal order, and we can link the individual Choi operators together via the link product. Using this within the labs and the mathematical definition of the link product (which reduces to $* = \otimes$ for operators with non-overlapping Hilbert spaces) to connect the process and the labs, we find for the statistics of the full protocol:
\begin{align}
	W  & \bigast_{j=1}^N \Big( \ket{\phi^+}\bra{\phi^+}_{P_j^I \tilde{P}_j^I} * \ket{\phi^+}\bra{\phi^+}_{P_j^O \tilde{P}_j^O} \Big) \bigast_{j \in P} \ket{\mathbb 1} \bra{\mathbb 1}_{\tilde{O}_j I_j^M}  \bigast_{j \in F} \ket{\mathbb 1}\bra{\mathbb 1}_{O^M_j \tilde{I}_j} \nonumber\\
	& \bigast_{j \in P}  \Big( \ket{\phi^+}\bra{\phi^+}_{I_j P_j^I} * (M_j)_{\tilde{P}^I_j \tilde{O}'_j} * (BSM_{m_j})_{\tilde{O}_j \tilde{O}'_j \tilde{P}_j^O} * (CU_{m_j})_{I_j^M P_j^O O_j} \Big) \nonumber \\
	&\bigast_{j \in F} \Big( (BSM_{m_j})_{O_j^M I_j P_j^I} * (CU_{m_j})_{\tilde{I}_j \tilde{P}^I_j \tilde{I}'_j} * (M_j)_{\tilde{I}'_j \tilde{O}_j} * \ket{\phi^+} \bra{\phi^+}_{\tilde{O}_j \tilde{P}_j^O} * \ket{\mathbb 1} \bra{\mathbb 1}_{P_j^O O_j} \Big) \label{Equation:PartialTeleportationLink}
\end{align}
The first line in Equation~\eqref{Equation:PartialTeleportationLink} is $W_{ext}$, the second line the instruments of the agents $j \in P$ and the third line the instruments of the agents $j \in F$.

Instead of calculating this convoluted expression, we use the associativity of the link product to factor out $W$. Then, the rest in Equation~\eqref{Equation:PartialTeleportationLink} is the same mathematical form that we would get if all of the parts of the teleportation protocol happened within a single lab with a definite causal structure. In particular, this mathematical form is independent of whether $W$ is a definite causal order, or an indefinite causal structure. Using the fact that the link product is the correct way to combine circuit fragments, and the teleportation protocol from the first section of the Supplementary Material, we therefore conclude that Equation~\eqref{Equation:PartialTeleportationLink} reduces to 
\begin{align}
	 \prod_{j\in P} \frac{1}{d^2_{j, \mathrm{in}}} \cdot \prod_{j \in F}\frac{1}{d^2_{j, \mathrm{out}} } \cdot\mathrm{Tr}[W^T \cdot \bigotimes_j M_j].
\end{align}
This means the teleportation behaves as if the instruments $M_j$ were inserted directly into $W$, up to the post-selection probabilities.

\end{document}